\documentstyle[12pt]{article}
\title{\vspace{-1.in} \hfill {\small\rm NTUA-78/00} \\~\\~\\
Strongly-interacting Fermions \\ from a higher-dimensional \\
Unified Gauge Theory}
\author{ George
Triantaphyllou\footnote{e-mail:georgetr$@$physics.ntua.gr}
 $\;$ and George
 Zoupanos\footnote{e-mail:George.Zoupanos$@$cern.ch}
$\;$\\~
\\{\it Physics Department, National Technical University,
Athens}\\ {\it Zografou Campus, 157 80 Athens, GREECE }}
\begin{document}
 \setlength{\baselineskip}{16pt}
 \maketitle
 \thispagestyle{empty}
\begin{abstract}
The method of coset-space dimensional reduction is employed in
order to proceed  from a gauged $E_{8} \times E_{8}^{\prime}$
unified theory defined in 10 dimensions to 4 dimensions.  The
resulting theory comprises the Standard Model along with a
strongly-interacting fermion sector which breaks the electroweak
symmetry dynamically at the right scale.
 \vspace{2.in}
\end{abstract}
\vfill
\setcounter{page}{0}
\pagebreak

\section{Introduction}

Efforts to describe the theory hidden behind not only the Higgs
mechanism but also the fermion-generation pattern has led to
various unified gauge theories. These typically contain
fundamental scalar particles with masses around the weak scale and
usually below 1 TeV. While a space-time global symmetry like
supersymmetry can stabilize scalar masses at such low scales, it
cannot explain why nature has chosen this particular value for the
weak scale.

 Since unified field theories use a gauge symmetry to
explain the hierarchy between the strong-interactions scale and
the unification scale $\Lambda_{GUT}$, it is reasonable to imagine
that the hierarchy between the weak scale and $\Lambda_{GUT}$ is
also due to a local symmetry. This leads one to replace the
perturbative Higgs sector with a non-perturbative effective one
providing a dynamical symmetry breaking mechanism based on
high-colour representations \cite{high} of ordinary QCD or on new
local symmetries like the ones introduced in technicolour
\cite{techni} or top-colour \cite{top} scenaria.

An effort to construct a unified theory which avoids fundamental
scalars near the weak scale and in parallel addresses the strong
CP problem appeared recently \cite{george1}. It introduces new
fermions with interchanged weak quantum-number assignments,
referred to as katoptrons, which, in contrast to the hitherto
known mirror fermions, interact strongly with each other under a
new ``horizontal" gauge interaction. While being unified with the
other SM interactions at a unification scale $\Lambda_{GUT}$
consistent not only with flavour-changing neutral-current and
proton-lifetime bounds but also with small SM-neutrino masses
\cite{george2}, the katoptron horizontal interaction becomes
naturally strong around the weak scale. Other dynamical symmetry
breaking models can hardly offer such a unification perspective
which tackles the hierarchy problem. This approach is also to be
distinguished from models using weak horizontal interactions
between SM fermions in attempts to understand family mixing and
weak CP violation \cite{hori}, since only the katoptrons transform
under the new generation symmetry.

 On the other hand, generating mass for the
SM fermions depends on the breaking of this new gauge symmetry,
which is denoted here by $SU(3)^{\prime}$  (and introduced under
the notation $SU(3)_{2G}$ in \cite{george2}). While fermion
composite operators which can break $SU(3)^{\prime}$ dynamically
exist in the theory \cite{george1}, it has still to be shown that
they assume the values needed to reproduce correctly the SM
fermion-mass spectrum.

The approach advocated here has some clear phenomenological
advantages over previous dynamical symmetry breaking models. The
$T$ (or alternatively the $\Delta\rho$) parameter is naturally
small, since the operators responsible for the splitting between
the top- and bottom-quark masses are $SU(2)_{L}\times U(1)_{Y}$
invariant, in contrast with technicolour for instance, where the
relevant  extended-technicolour operators violate the electroweak
symmetry. Moreover, within the present framework the $S$ parameter
can also be kept small, mainly for three reasons. First, the
strong gauge group is broken, second Majorana katoptron-neutrinos
contribute negatively to $S$, and third, vertex corrections can
have a large negative effect on this parameter. Interestingly
enough, these vertex corrections are consistent with the large
deviation of the experimentally measured weak right-handed
bottom-quark coupling $g_{R}^{b}$ from its SM value
\cite{george3}.  This is in contrast with technicolour theories
for instance, which can hardly provide a cancellation mechanism
for the $S$ parameter, since there are no sizable contributions to
$g_{R}^{b}$ and the technicolour group is left intact.

The purpose of the present study is to show explicitly how such a
model could result from a more fundamental theory. In particular,
the 10-dimensional $E_{8} \times E_{8}^{\prime}$ model is one of
the very few to possess the advantage of anomaly freedom
\cite{Schwartz} and is extensively used in efforts to describe
quantum gravity along with the observed low-energy interactions in
the heterotic-string framework \cite{GSW}. It will be therefore
used in the following as a starting point for our investigation.
The results obtained indicate that it constitutes a very solid
basis indeed for the understanding of our world.

\section{Going from 10 to 4 dimensions}

\subsection{The use of coset spaces}

As promised above, we start with the gauge group $G=E_{8} \times
E_{8}^{\prime}$ defined in 10 space-time dimensions. The
exceptional Lie group $E_{8}$ has the unique property of having
its 248-dimensional fundamental and adjoint representations
identical. Therefore, spin-1/2 and spin-1 particles are both taken
to transform like ${\bf (248,1)}$ and ${\bf (1,248)}$ under the
above group structure, and the theory is at this level
supersymmetric and anomaly-free. The spin-1/2 fields are taken to
be Weyl-Majorana fermions. By use of the 10-dimensional chirality
operator, the ${\bf (248,1)}$ and ${\bf (1,248)}$ fermion
representations are defined to be left- and right-handed
respectively.

Making connection with our 4-dimensional world leads one to
consider 10-dimensional space-times of the form $M^{4}\times B$,
where $M^{4}$ is the usual Minkowski space and $B$ is a
6-dimensional compactified manifold the structure of which has to
be determined. Coset-space dimensional reduction (CSDR) of
higher-dimensional gauge theories \cite{Forg1}-\cite{Zoup}
provides a very elegant method of analyzing in detail the
resulting 4-dimensional models.

In the CSDR approach, the manifold $B$ is taken to be a coset
space $S/R$, where $S$ and $R$ are compact Lie groups and
\begin{equation}
{\rm dim}(B) = {\rm dim}(S) - {\rm dim}(R).
\end{equation}
The group $S$ can be considered as an $R$-bundle over $B$. The
group $R$ is taken to be a subgroup of both $S$ and $G$. The fact
that $R$ is not trivial is a necessary condition for a non-trivial
topology for the manifold $B$, something which is needed for the
survival of chiral fermion fields in 4 dimensions. The fact that
the 4-dimensional Lagrangian is independent of the extra
coordinates is then guaranteed by gauge invariance.

Embedding the symmetry $R$ in $G$ gives an interesting geometrical
content to some of the gauge symmetries of the theory. Dimensional
reduction from 10 to 4 dimensions is thus accompanied with rank
reduction resulting to a surviving gauge symmetry $H \subset G$,
which is the centralizer $C_{G}(R)$ of the group $R$ in $G$.
Furthermore, the extra compactified dimensions used in CSDR offer
a natural framework for the unification of gauge and scalar
fields. The latter have interaction potentials which can lead to
spontaneous symmetry breaking, leading us from a unified gauge
theory to the SM.

Recently, higher-dimensional theories were considered beyond the
classical level \cite{Kubo} and were given a quantum meaning in
the sense of the Wilson renormalization group in agreement with
the treatment involving massive Kaluza-Klein excitations
\cite{Vene}. The CSDR approach can therefore be exploited in the
study of higher-dimensional unified quantum field theories
independently of more general frameworks like string theory.

Very strict rules \cite{Forg1}, \cite{Manton}, \cite{Slan}
determine which fields finally survive, i.e. remain massless,
after this process, since gauge transformations have to be
compensated by the action of the symmetry group $S$. These rules
are a guiding light for model-building, ruling out groups that
lead to unacceptable phenomenologies. It will be seen for instance
that katoptrons surviving at low energies can be obtained only by
coset spaces which are non-symmetric \cite{nonsym}. In connection
with initial $E_{8}$ groups, these lead interestingly enough to an
$E_{6}$ unification group.

In particular, one has to decompose the adjoint representations of
the groups $G$ and $S$  under $R \times H$ and $R$ respectively
according to
\begin{eqnarray}
{\rm adj}(G) &=& ({\rm adj}(R), 1) + (1, {\rm adj}(H)) + \sum_{i}
(r_{i}, h_{i}) \nonumber \\ {\rm adj}(S) &=& {\rm adj}(R) +
\sum_{i} s_{i}.
\end{eqnarray}

\noindent The only spin-1 fields surviving are the ones
transforming under the adjoint representation of $H$. The spin-0
fields that appear after dimensional reduction, even though
initially absent, are the ones transforming like $h_{i}$ under
$H$, and only for those $i$'s for which $r_{i}=s_{i}$.

As regards spin-1/2 fields, one decomposes the fermion
representation $F$ of $G$ and the spinor representation of $SO(6)$
under $R\times H$ and $R$ respectively according to
\begin{eqnarray}
F &=& \sum_{i}(r_{i},h_{i}) \nonumber \\ \sigma &=& \sum_{i}
\sigma_{i}.
\end{eqnarray}

\noindent The only fermion fields surviving are the ones
transforming like $h_{i}$ under $H$, and only for those $i$'s for
which $r_{i}=\sigma_{i}$.

When studying $E_{8} \times E_{8}^{\prime}$ models \cite{Suema},
it is customary to identify the origin of the fields transforming
under $E_{8}^{\prime}$ with some obscure ``hidden" or ``shadow"
world that interacts only gravitationally with ours. The
philosophy here is different, because katoptron fermions originate
from this new world. Since these fermions will finally assume the
role of a dynamical Higgs sector, they should have quantum-number
assignments similar (but not identical) to the ones of their SM
partners.

A way to achieve this goal is to make use of a discrete abelian
subgroup of $G$ consisting of two elements, which we denote by
$Z^{E_{6}}_{2}$. The action of its non-trivial group element
corresponds to an outer automorphism that interchanges the $E_{6}$
subgroups of the two $E_{8}$'s. (Analogously, $Z^{E_{8}}_{2}$
interchanges the two $E_{8}$'s.) Making use of $Z_{2}^{E_{6}}$ has
the effect of reducing further the rank of the surviving symmetry
$H$ in a manner analogous to the construction in Ref.\cite{Forg2},
as will be seen shortly. In the following, a particular
6-dimensional non-symmetric coset space is analyzed and shown to
lead to an acceptable phenomenology.

\subsection {CSDR with $S=Sp(4)$, $R=(SU(2) \times U(1))_{{\rm non-max}}$}

We consider a Lie group $R=SU(2)\times U(1)$ embedded
non-maximally into $S=Sp(4)$ and into $E_{8} \subset G$, i.e. into
the exceptional group under which the Weyl-Majorana fermions of
the model are left-handed. The Euler characteristic of
$Sp(4)/(SU(2)\times U(1))_{{\rm non-max}}$ is equal to $\chi=4$,
and a priori the number of copies of the fermion representations
is, according to the index theorem, equal to $|\chi/2|=2$.

Compactifying on $B_{0}=S/R$ leads to the following decompositions
of the adjoint and spinor representations of $SO(6)$ and $Sp(4)$
under $R$ respectively:
\begin{eqnarray}
SO(6) \supset (SU(2) \times U(1))_{{\rm non-max}},& {\bf 4} =
({\bf 1},0)+({\bf 1},2) + ({\bf 2},-1) \nonumber
\\ &{\bf {\bar 4}} = ({\bf 1},0)+({\bf 1},-2) + ({\bf 2},1)
 \nonumber \\ Sp(4) \supset (SU(2) \times U(1))_{{\rm non-max}},& {\bf 10} =
({\bf 1+3},0)+({\bf 2},\pm 1) + ({\bf 1}, \pm 2)
\end{eqnarray}

\noindent The adjoint representation of $E_{8}$ decomposes under
$(SU(2) \times U(1))_{{\rm non-max}} \times E_{6}$ as follows:
\begin{eqnarray}
{\bf 248}&=&({\bf 1},0,{\bf 78})+({\bf 1+3},0,{\bf 1})+({\bf
2},\pm 3,{\bf 1}) \nonumber
\\ &+&({\bf 1},-2,{\bf 27})+({\bf 2},1,{\bf 27})
+({\bf 1},2,{\bf \overline{27}})+({\bf 2},-1,{\bf \overline{27}})
\end{eqnarray}

 These decompositions are not altered if the compactification is performed on
 the space $B=(S/R)\times (Z_{2}^{E_{8}}/Z_{2}^{E_{6}})$, i.e. when $R$ is
 replaced by ${\tilde R} \equiv R \times
Z_{2}^{E_{6}}$ and the corresponding fields are taken to be
$Z^{E_{6}}_{2}$ singlets, with $Z^{E_{6}}_{2}$ defined as
previously. Since $C_{E_{8}}(R)= E_{6} \;\left(\times
Z^{R}_{2}\times U(1)\right)$, with $Z^{R}_{2}\times U(1)$ the
center of $SU(2)\times U(1)$ (the superscripts of the various
$Z_{2}$ symmetries in this paper have each obviously different
meaning), and $C_{E_{8}^{\prime}}(E_{6}^{\prime})=SU(3)^{\prime}$,
the centralizer $C_{G}({{\tilde R}})$ is equal to
 \begin{equation}
 H = E_{6}^{D}  \times SU(3)^{\prime} \;
  \left( \times Z^{R}_{2}\times U(1)\right),
\end{equation}

\noindent  where $E_{6}^{D}$ is the diagonal subgroup of the
$E_{6}$ subgroups of the two $E_{8}$'s.

That $H$ given above is indeed the surviving gauge symmetry after
compactification can be checked explicitly  by enumerating the
spin-1 degrees of freedom which are left invariant by the action
of ${\tilde R}$. In the absence of the $Z_{2}^{E_{6}}$ symmetry,
$H$ would have been given by $E_{6}  \times E_{8}^{\prime}\;
\left(\times Z^{R}_{2}\times U(1)\right)$. The role of
$Z_{2}^{E_{6}}$ is to keep only the diagonal subgroup of
$E_{6}\times E_{6}^{\prime} \subset E_{8}\times E_{8}^{\prime}$
unbroken, eliminating all skew-symmetric contributions.
Physically, it renders the compactification process more symmetric
with respect to the two $E_{8}$'s.

Furthermore, the SM-fermion quantum numbers under $Z^{R}_{2}\times
U(1)$, the center of $SU(2)\times U(1)$, are equal to $(1,-2),
(1,1,)$ and $(-1,1)$. The center survives after CSDR, and it is
identified in the following with the family symmetry of the SM
which differentiates between the three SM generations. This
symmetry is taken to be global and the $U(1)$ coupling is
accordingly switched-off in order to avoid problems with
flavour-changing neutral currents at lower energies.

Under the gauge structure $H=E_{6}^{D}  \times SU(3)^{\prime} \;
\left( \times Z^{R}_{2} \times U(1)\right)$ defined above, the
CSDR rules give the following surviving 4-dimensional fields:
\begin{eqnarray}
{\rm spin-1/2:} &({\bf 27,3}) &{\rm Katoptrons} \nonumber
\\
         & ({\bf  78,1})+({\bf 1,8}) &{\rm Vector \; fermions}
         \nonumber \\
         & 2\times ({\bf 27,1})_{a} &{\rm Include\;SM\; fermions\;}
         \nonumber\\
{\rm spin-0:}   & 2\times({\bf 27,1})_{a} &{\rm Higgs \;sector\;}
\end{eqnarray}

\noindent where the fields transforming under $E^{D}_{6}$ as ${\bf
27}$ and ${\bf \overline{27}}$ are identified by the Majorana
condition, and the subscript $a=1,2,3$ serves as a generation
index corresponding to $Z^{R}_{2}\times U(1)$ . In the above, we
also indicate in which sector fields which are known to us from
the SM and katoptron model are contained. The vector fermions
transforming like $({\bf 1,8})$ and $({\bf 78,1})$ are not
protected by any gauge symmetry, so according to the general
argumentation on the survival hypothesis they acquire large
gauge-invariant masses of the order of the compactification scale
and disappear from the low-energy spectrum. The torsion of the
non-symmetric space $B$ \cite{Kapet} is taken to be such that
katoptrons remain massless.

The present coset space admits two different scales \cite{Castel},
something that could be useful in the subsequent breaking of $H$,
as will be discussed later. Furthermore, it can be checked that
this breaking leads to an anomaly-free 4-dimensional theory, since
it satisfies the equation
\begin{equation}
l(G)=60,
\end{equation}

\noindent where $l(G)$ is the sum of the indices of all the
representations of $R$ appearing when the adj($G$) representation
is decomposed under $R \times H$ \cite{Lust2}. After having
analyzed the geometrical rank reduction of $G$ down to $H$, one
has to study the subsequent breaking of $H$ down to the SM.

\section{Symmetry breaking to the Standard Model}

\subsection{Breaking by Wilson lines}

The simply-connected group $S=Sp(4)$ considered has a $Z_{2}$
symmetry as center (recall that $Sp(4)/Z_{2} \approx SO(5)$), and
this can be employed here to serve in a gauge-symmetry breaking
mechanism by Wilson lines \cite{Hoso}. The embedding of this
abelian discrete symmetry in the $SU(2)_{L}$ subgroup of
$E_{6}^{D}$, which we denote by $Z_{2} \equiv Z_{2}^{SU(2)_{L}}$,
can be defined via the following homomorphism involving its
non-trivial group-element $g$ \cite{Aguila}:
\begin{equation}
Z_{2}^{SU(2)_{L}} \ni g \longrightarrow U_{g} = {\bf 1} \otimes
{\bf 1} \otimes \left( \begin{array}{ccc}1&&\\&-1&\\&&-1
\end{array} \right) \in SU(3)_{C} \times SU(3)_{R} \times
SU(3)_{L}
\end{equation}

\noindent where $SU(3)^{\otimes 3}$ is a maximal subgroup of
$E_{6}^{D}$, the second and third row of the (diagonal)
$SU(3)_{L}$ factor correspond to $SU(2)_{L}$,  and of course
$U_{g}^{2}=1$.

We then consider dimensional reduction over the coset space
\begin{equation}
{\tilde B}=\left(S/(R \times Z_{2}^{SU(2)_{L}})\right )\times
(Z_{2}^{E_{8}}/Z_{2}^{E_{6}}),
\end{equation}

\noindent The original gauge group $G$ is broken at the
compactification scale, which is identified here with the
gauge-coupling unification scale $\Lambda_{GUT}$, down to
\begin{equation}
H^{\prime} = SU(6) \times SU(2)_{L}\times SU(3)^{\prime}\;
\left(\times Z^{R}_{2}\times U(1)\right),
\end{equation}

\noindent where $SU(6)\times SU(2)_{L} \subset E_{6}^{D}$. The
original $|\chi/2|=2$ copies of fermion and scalar fields are then
further reduced to a single copy due to the action of
$Z_{2}^{SU(2)_{L}}$. Obviously, one has to distinguish the
topological role of this symmetry from the role of $Z^{R}_{2}$
which merely differentiates the fermion families via quantum
numbers.

The effect of $Z_{2}^{SU(2)_{L}}$ is apparently at the heart of
the parity asymmetry of our world.  One of its side-effects is to
break the original supersymmetry at the compactification scale,
since fermion fields lose some of their bosonic partners. It is
reminded here that the present model does not need low-energy
supersymmetry, since the hierarchy problem is solved by the gauge
symmetry $SU(3)^{\prime}$ \cite{george2}.

\subsection{Further breaking by a Higgs mechanism}

 One of the scenarios presented in \cite{Aguila} is subsequently
realized. The Higgs fields transform under $SU(6) \times
SU(2)_{L}$ like ${\bf (6,2) + (15,1)}$. Only the ${\bf (15,1)}$
Higgses which are invariant under $Z_{2}^{SU(2)_{L}}$ remain
light, and one of their copies is taken to develop a non-zero
vacuum expectation value at the compactification scale. This
breaks spontaneously the gauge symmetry further down to
\begin{equation}
H^{\prime \prime} = SU(4)_{PS} \times SU(2)_{R} \times SU(2)_{L}
 \times SU(3)^{\prime}\;\left(\times Z^{R}_{2}\times U(1)\right),
\end{equation}

\noindent where $SU(4)_{PS}$ is the usual Pati-Salam symmetry. The
${\bf 27}$ representation of $E_{6}$ decomposes under
$SU(4)_{PS}\times SU(2)_{R} \times SU(2)_{L}$ like
\begin{equation} {\bf 27} = ({\bf {\bar 4},1,2})+({\bf 1,2,2})+({\bf
4,2,1})+({\bf {\bar 4},1,1})+({\bf 1,2,1})+({\bf 1,1,1})
\end{equation}

 The remaining generations
of Higgses transform under $SU(4)_{PS} \times SU(2)_{R}$ like
${\bf (4,2)+({\bar 4},1)+(1,2)+(1,1)}$. The ${\bf (4,2)}$ Higgs
field is subsequently taken to acquire a non-zero vacuum
expectation value and break spontaneously the gauge symmetry
$SU(4)_{PS}\times SU(2)_{R}$ further down to $SU(3)_{C}\times
U(1)_{Y}$ at the Pati-Salam scale $\Lambda_{PS}$, giving the final
symmetry
\begin{equation}
H^{\prime \prime \prime } = SU(3)_{C} \times SU(2)_{L} \times
U(1)_{Y}  \times SU(3)^{\prime}\;\left(\times Z^{R}_{2}\times
U(1)\right),
\end{equation}

\noindent which includes the familiar SM groups. The fact that the
coset space considered in the last section admits two different
scales could be at the origin of the
 relatively small hierarchy between the Pati-Salam symmetry breaking scale
$\Lambda_{PS}$ and the unification scale $\Lambda_{GUT}$
\cite{george2}.

 There is no surviving symmetry preventing scalar particles
from obtaining large masses after these breakings. Spin-1/2
particles remain light only if they are chiral, the others gaining
compactification-scale masses. One then recovers the gauge and
matter content which is the starting point of \cite{george1}, by
taking the fermions in the representations of the original $E_{8}$
and $E_{8}^{\prime}$ groups to be left-handed and right-handed
respectively, as stated in the beginning.

Therefore, one reproduces the three SM generations and a katoptron
generation which has interchanged left-right $SU(2)_{L}\times
SU(2)_{R}$ quantum numbers and transforms in addition in the
fundamental representation of the gauge group $SU(3)^{\prime}$.
Under $H^{\prime \prime \prime}$, the fields transform like
\begin{eqnarray}
&{\rm SM\;fermions} &{\rm Katoptrons} \nonumber
\\
&q_{L}:({\bf 3,\;2},\;1/3,\;{\bf 1})_{a}  & q_{R}^{K}:({\bf
3,\;2},\;1/3,\;{\bf 3}) \nonumber \\ &l_{L}:({\bf
1,\;2},\;-1,\;{\bf 1})_{a}  &l_{R}^{K}:({\bf 1,\;2},\;-1,\;{\bf
3})\nonumber \\&q_{R}^{c}:({\bf {\bar
3},\;1},\;~^{-4/3}_{+2/3},\;{\bf 1})_{a}  & q_{L}^{K\;c}: ({\bf
{\bar 3},\;1},\;~^{-4/3}_{+2/3},\;{\bf 3}) \nonumber
\\&l_{R}^{c}:
 ({\bf 1,\;1},\;~^{0}_{2},\;{\bf 1})_{a} & l_{L}^{K\;c}:
 ({\bf 1,\;1},\;~^{0}_{2},\;{\bf 3}),
   \nonumber \\
\end{eqnarray}

\noindent where the superscript $K$ denotes katoptron fields, $c$
charge conjugation, the subscripts $L$ and $R$ left- and
right-handed fields, and $q$ and $l$ quark and lepton fields
respectively.
 The group $SU(3)^{\prime}$ is
asymptotically free and provides the mechanism responsible for the
dynamical breaking of the electroweak symmetry at the right scale
via katoptron condensates.

\section{Discussion}

 Starting with a higher-dimensional gauge
field theory, we presented an effort to produce a picture
consistent with current phenomenology and which in addition
includes a dynamical Higgs sector.
 The need to obtain eventually the SM group structure
 at lower energies in 4 dimensions places severe constrains on
 the compactification manifolds considered.
The gauge-symmetry-breaking sequence of \cite{george2} can be
reproduced by use of Wilson lines for example, if the group
manifold $S$ has $Z_{2}$ as center, so $S=Sp(4)$ is left as a
unique choice (for $S$ semisimple) leading to a 6-dimensional
non-symmetric manifold $B$.

 Moreover, in order to make connection with the unification picture
presented in \cite{george2}, one has to note that the $E_{6}$
group with three generations seems to be favored over $SO(10)$
with 4 generations considered in that reference as a unification
symmetry. In all other respects, the results and conclusions of
\cite{george2} remain unaltered, since the scenario with 4
generations was rejected there for other reasons.


The likelihood of the scenario presented here should be tested not
only for its theoretical consistency but also for its
phenomenological relevance in forthcoming experiments
\cite{george3}. It would then constitute one interesting example
trying to connect the abstract mathematical world of the Planck
scale with the experimental physical reality of collider data.

\noindent {\bf Acknowledgements} \\ We thank P. Forgacs, A.
Kehagias, D. L\"{ust} and D. Suematsu for interesting discussions
and for reading the manuscript. G.T. thanks the National Technical
University of Athens for their warm hospitality and support during
the months when this work was completed. G.Z. is partially
supported by the E.C. project ERBFMRXCT960090.


\begin{thebibliography}{99}
\bibitem{high} W. Marciano, {\it Phys. Rev.} {\bf D 21} (1980),
315; K. Konishi and R. Tripiccione, {\it Phys. Lett.} {\bf B 121}
(1983), 403; G. Zoupanos, {\it Phys. Lett.} {\bf B 129} (1983),
315; D. L\"{u}st {\it et al.}, {\it Phys. Lett.} {\bf B 158}
(1985), 55; {\it Nucl. Phys.} {\bf B 268} (1986), 49; W. Bardeen,
Proceedings of {\it Rencontres de Moriond} (1986), p.207.
\bibitem{techni} For an early review, see E. Fahri and L.
Susskind, {\it Phys. Rep.} {\bf 74} (1981), 277.
\bibitem{top} C.T. Hill, {\it Perspectives of strong-coupling gauge
theories}, Nagoya Proceedings (1996), p. 54, and references
therein; {\it ibid.}, {\it Phys. Lett.} {\bf B 266} (1991), 419;
{\it ibid., } {\bf B 345} (1995), 483.
\bibitem{george1} G. Triantaphyllou, {\it J. Phys.} {\bf G 26}
(2000), 99.
\bibitem{george2} G. Triantaphyllou, {\it Eur. Phys. J.} {\bf C
10} (1999), 703.
\bibitem{hori} T. Maehara and T. Yanagida, {\it Prog. Theor.
Phys.} {\bf 60} (1978), 822; {\it ibid, } {\bf 61} (1979), 1434;
F. Wilczek and A. Zee, {\it Phys. Rev. Lett.} {\bf 49} (1979),
421; A. Davidson {\it et al., Phys. Rev. Lett.} {\bf 43} (1979),
92; G. Zoupanos, {\it Z. Phys.} {\bf C 11} (1981), 27; {\it ibid.,
Phys. Lett.} {\bf B 115} (1982), 221; M.B. Gavela and H. Georgi,
{\it Phys. Lett.} {\bf B 119} (1982), 141; E. Papantonopoulos and
G. Zoupanos, {\it Phys. Lett.} {\bf B 110} (1982), 465; {\it
ibid., Z. Phys.} {\bf C 16} (1983), 261; R. Decker {\it et al.,
Phys. Lett.} {\bf B 137} (1984), 83.
\bibitem{george3} G. Triantaphyllou, {\it Int. J. Mod. Phys.} {\bf A 15}
(2000), 265.
\bibitem{Schwartz} M.B. Green, J.H. Schwartz, {\it Phys. Lett.}
{\bf B 149} (1984), 117; L. Alvarez-Gaume and E. Witten, {\it
Nucl. Phys.} {\bf B 234} (1983), 269.
\bibitem{GSW} See for instance
M.B. Green, J.H. Schwarz and E. Witten, Superstring Theory,
Cambridge University Press (1987); D. L\"{u}st and S. Theisen,
{\it Lectures on String Theory}, Lecture Notes in Physics, {\bf
Vol. 346}, Springer Verlag, Heidelberg (1989).
\bibitem{Forg1} P. Forgacs and N.S. Manton, {\it Commun. Math.
Phys.} {\bf 72} (1980), 15; E. Witten, {\it Phys. Rev. Lett.} {\bf
38} (1977), 121.
\bibitem{Zoup} For a review see D. Kapetanakis and G. Zoupanos,
{\it Phys. Rep.} {\bf 219} (1992), 1, and references therein. See
also Y.A. Kubyshin {\it et al.}, {\it Dimensional Reduction of
Gauge Theories, Spontaneous Compactification and Model Building},
 Lecture Notes in Physics, {\bf Vol. 349}, Springer Verlag,
 Heidelberg (1989); F.A. Bais {\it et al., Nucl. Phys.} {\bf B
 263} (1986), 557.
\bibitem{Kubo} T. Kobayashi, J. Kubo, M. Mondragon and G.
Zoupanos, {\it Nucl. Phys.} {\bf B 550} (1999), 99; J. Kubo, H.
Terao and G. Zoupanos, {\it Nucl. Phys.} {\bf B 574} (2000), 495;
S. Ejiri, J. Kubo and M. Murata, hep-ph/0006217.
\bibitem{Vene} T.R. Taylor and G. Veneziano, {\it Phys. Lett.}
{\bf B 212} (1988), 147; K. Dienes, E. Dudas and T. Gherghetta,
{\it Phys. Lett.} {\bf B 436} (1998), 311; {\it ibid.,} {\it Nucl.
Phys.} {\bf B 537} (1999), 47.
\bibitem{Manton} N.S. Manton, {\it Nucl. Phys.} {\bf B 193}
(1981), 502.
\bibitem{Slan} C. Chapline and R. Slansky, {\it Nucl. Phys.} {\bf B
209} (1982), 461.
\bibitem{nonsym} P. Forgacs, Z. Horvath and L. Palla, {\it Z.
Phys.} {\bf C 30} (1986), 261.
\bibitem{Suema} T. Matsuoka and D. Suematsu, {\it Nucl. Phys.}
{\bf B 274} (1986), 106; {\it ibid., Prog. Th. Phys.}, {\bf 76}
(1986), 886.
\bibitem{Forg2} P. Forgacs {\it et. al., Nucl. Phys.} {\bf B
308} (1988), 477; for an alternative way to obtain a diagonal
subgroup, see P. Forgacs and G. Zoupanos, {\it Phys. Lett.} {\bf B
148} (1984), 99.
\bibitem{Kapet} See for instance
D. Kapetanakis and G. Zoupanos, {\it Z. Phys.} {\bf C 56} (1992),
91.
\bibitem{Castel} K. Farakos {\it et al., Nucl. Phys.} {\bf B 291}
(1987), 128; {\it ibid., Phys. Lett.} {\bf B 191} (1987), 135; F.
Muller-Hoissen and R. Stuckl, {\it Class. Quant. Grav.} {\bf 5}
(1988), 27; L. Castellani, L.J. Romans and N.P. Warner, {\it Ann.
Phys.} {\bf 157} (1984), 394.
\bibitem{Lust2} D. L\"{u}st, {\it Nucl. Phys.} {\bf B 276} (1986),
220; E. Witten, {\it Phys. Lett.} {\bf B 149} (1984), 351; K.
Pilch and A.N. Schellekens, {\it Nucl. Phys.} {\bf B 259} (1985),
637.
\bibitem{Hoso} Y. Hosotani, {\it Phys. Lett.} {\bf B 126} (1983),
193; G. Zoupanos, {\it Phys. Lett.} {\bf B 201} (1988), 301; N.G.
Kozimirov, V.A. Kuzmin and I.I. Tkachev, {\it Sov. J. Nucl. Phys.}
{\bf 49} (1989), 164; D. Kapetanakis and G. Zoupanos, {\it Phys.
lett.} {\bf B 232} (1989) 104.
\bibitem{Aguila} F. del Aguila {\it et. al., Nucl. Phys.} {\bf B 272}
(1986), 413.
\end{thebibliography}
\end{document}